\documentclass[nofootinbib, twocolumn, aps, prd, superscriptaddress, showkeys]{revtex4-2}
\usepackage{mathtools}
\usepackage{mathrsfs}
\usepackage{amsfonts}
\usepackage[scale=1.05]{newtxtext}
\usepackage[scale=1.05]{newtxmath}
\usepackage[colorlinks=true, citecolor=blue, linkcolor=blue, urlcolor=blue!50!black]{hyperref}
\usepackage{booktabs}
\usepackage{enumitem}
\usepackage{graphicx}
\usepackage{orcidlink}

\newcommand{\dif}{\ensuremath{\mathrm{d}}}
\newcommand{\e}{\ensuremath{\mathrm{e}}}
\newcommand{\ii}{\ensuremath{\mathrm{i}}}
\newcommand{\RR}{\ensuremath{\mathbb{R}}}
\newcommand{\CC}{\ensuremath{\mathbb{C}}}
\newcommand{\real}{\ensuremath{\text{Re}}}

\newcommand{\cc}[1]{\overline{#1}}
\newcommand{\Kl}{\ensuremath{{K^\mathrm{l}}}}
\newcommand{\Ks}{\ensuremath{{K^\mathrm{s}}}}
\newcommand{\Kv}{\ensuremath{K^\mathrm{v}}}
\newcommand{\Ktt}{\ensuremath{K^\mathrm{tt}}}

\begin{document}

\title{Gravitational wave polarization modes and the kinematical tensors in general relativity and beyond}

\author{Cynthia Maldonado\,\orcidlink{0009-0007-5519-6804}
}
\email{cyn@ciencias.unam.mx}
\affiliation{Departamento de F\'{\i}sica, Facultad de Ciencias, \href{https://ror.org/01tmp8f25}{Universidad Nacional Autónoma de México}, Ciudad de M\'exico, 04510, M\'exico.}

\author{Francisco Nettel\,\orcidlink{0000-0001-7372-987X}}
\email{fnettel@ciencias.unam.mx}
\affiliation{Departamento de F\'{\i}sica, Facultad de Ciencias, \href{https://ror.org/01tmp8f25}{Universidad Nacional Autónoma de México}, Ciudad de M\'exico, 04510, M\'exico.}

\author{Pedro A. S\'anchez\,\orcidlink{0009-0008-7437-0561}}
\email{pedro.sanchez.s@cinvestav.mx}
\affiliation{Departamento de F\'{\i}sica, \href{https://ror.org/009eqmr18}{Centro de Investigaci\'on y de Estudios Avanzados del Instituto Politécnico Nacional}, Ciudad de M\'exico, 07360, M\'exico.}

\begin{abstract}
Relations between the kinematical tensors (the expansion, the shear, and the vorticity) and the polarization modes of gravitational waves are studied within the context of metric theories of gravity by considering freely falling test particles. After analyzing exact relations, we consider slowly moving particles under the influence of a weak gravitational field. Linearized plane waves of theories representative of those determined by a general second-order Lagrangian, including General Relativity, are shown to exemplify the following interconnections: between the transverse components of the shear and the transverse tensor polarization mode; between the expansion, and both the transverse scalar and the longitudinal polarization modes; and between the longitudinal-transverse components of both the shear and the vorticity, and the vector polarization mode. These relations, from the theoretical point of view, offer a novel insight into the interpretation of the polarization modes in terms of the kinematical tensors. Additionally, the inclusion of the kinematical tensors in the study of gravitational waves might provide a new outlook on their phenomenology.
\end{abstract}

\maketitle

\section{Introduction} \label{sec:intro}

At the heart of General Relativity (GR) and other metric theories of gravity, lies Einstein's equivalence principle (cf., e.g., chapter 2 of \cite{will2018theory}). It establishes an equivalence between the outcomes of local experiments, thus making nonlocal experiments essential to probe the gravitational field. Simple but physically rich nonlocal experiments can be conceived with a cloud of freely falling test particles, for instance by studying the kinematics and dynamics of their relative separation. Within a metric theory of gravity, neighboring particles with separation vector $\xi$ are predicted to have a relative velocity
\begin{equation} \label{eq:relative-velocity}
\frac{D \xi^a}{\dif\tau} = \frac{\theta}{3} \xi^a + {\sigma_{b}}^a \xi^b + {\omega_b}^a \xi^b,
\end{equation}
where $\tau$ is the proper time of the particles; and $\theta$, $\sigma$, and $\omega$ are their expansion, shear, and vorticity, respectively. Moreover, these neighboring particles are predicted to experience a relative, or tidal, acceleration given by the equation of geodesic deviation
\begin{equation} \label{eq:relative-acceleration}
\frac{D^2 \xi^a}{\dif\tau^2} = {K_b}^a \xi^b,
\end{equation}
written here in terms of the tidal tensor $K$. Then, measurements of the relative velocity between the particles reveal information about the kinematical tensors, namely $\theta$, $\sigma$, and $\omega$, from which some aspects about the gravitational field could be inferred. Similarly, measuring the relative acceleration reveals information about the part of the Riemann tensor encoded in the tidal tensor---its so-called electric part.

Some gravitational fields have a distinctive tidal tensor, as the following examples arising in GR illustrate (cf., e.g., \cite{szekeres1965gravitational}). A cloud of particles falling radially in a Schwarzschild spacetime, experiences a tidal acceleration that tends to elongate it in the radial direction and compress in the transverse directions. On the other hand, a cluster of galaxies in a spatially homogeneous and isotropic universe, tends to expand accelerating equally in all directions. Contrastingly, a cloud of test particles is affected by a plane gravitational wave, such that only particles separated in directions transverse to the wave spatial propagation direction accelerate while preserving the transverse area of the cloud.

For a general gravitational field, the six independent components of the tidal tensor have different effects on a cloud of test particles. This is particularly relevant to detect a gravitational wave. Given a wave with definite propagation direction, the tidal tensor is uniquely decomposed into its purely longitudinal, mixed longitudinal-transverse, and purely transverse parts, the latter being further decomposed into its tensor and scalar parts. These constitute the polarization modes of the wave~\cite{eardley1973gravitational, will2018theory}. GR predicts that a vacuum propagating wave has only the two polarization modes of the transverse tensor part, but many other metric theories of gravity allow more polarization modes (cf. \S 11.4.2 of \cite{will2018theory}, and \cite{hou2018polarizations, capozziello2019gravitational, schumacher2023gravitational, dong2024polarization, lai2024polarization}). The search for observational evidence of polarization modes beyond GR, is actively pursued by different collaborations (cf. \cite{abbott2018search, arzoumanian2021nanograv, agazie2024nanograv, abbott2025tests} together with \S 4.3.7 of \cite{yunes2025gravitational} and references cited therein).

Likewise, the study of the kinematical tensors has its own relevance for some types of gravitational fields, standing out cosmological models for which it has been a key element in their construction (cf. \S 5.1 of \cite{ellis1971relativistic,ellis2009republication} and \S 9.1 of \cite{ellis2012relativistic}). However, it is not so common in the description of gravitational waves, with notable exceptions such as~\cite{herrera2000influence, bishop2022effect} and to some extent \cite{bini2000gyroscopes, seraj2022precession}.
Then, the use of a complementary description of the effects of
gravitational waves on test particles, such as the one provided by the kinematical tensors, remains largely unexplored and might prove useful.

Following this line of reasoning, in this work we propose studying the kinematical tensors in the context of gravitational waves of metric theories of gravity, how they relate to the polarization modes, and use them to compare some generic modified theories of gravity with GR. Intuition leads one to expect that the kinematical tensors of a cloud of test particles subject to the effects of a gravitational wave are related to the polarization modes of the wave. Concretely, one expects, at least, the following relations: between the transverse components of the shear and the transverse tensor polarization modes; between the longitudinal component of the shear and the longitudinal polarization mode; and between the expansion and the transverse scalar polarization mode. Specifically, we analyze these relations for gravitational waves with a definite propagation direction, within the context of GR and other metric theories of gravity, in both the strong and weak field regimes. To our knowledge, the full set of kinematical tensors had not been used to study gravitational waves.

This paper is organized as follows. In \S \ref{sec:general-field} we review the definitions and the physical meaning of the kinematical tensors, the tidal tensor, and the polarization modes for a cloud of freely falling test particles. Additionally, we express the polarization modes in terms of the kinematical tensors, and discuss what the kinematical tensors offer to the analysis of gravitational waves. Then, in \S \ref{sec:weak-waves} we specialize the analysis to slowly moving particles in a weak gravitational field, obtaining expressions for both the kinematical tensors and the polarization modes at the considered order. Next, in \S \ref{sec:examples} we evaluate the obtained expressions on linearized gravitational plane waves of a selection of metric theories representative of those determined by a Lagrangian that is a general function of the Riemann tensor, and compare them with GR. In appendix~\ref{app:polarization-modes-NP} we write exact expressions for the polarization modes in terms the irreducible parts of the Riemann tensor, both in terms of real-valued components and in terms of the Newman-Penrose curvature scalars.

The \emph{conventions} used in this work are the following. The mostly plus signature is used. Spacetime indices are denoted by $a, b, c, d, e, f$; while spatial indices are denoted by $i, j, k, l, m$. The summation convention will be adopted for pairs of repeated indices, both spacetime and spatial.
The Riemann tensor is defined such that Ricci identities yield ${R^a}_{bcd} X^b = 2 \nabla_{[c} \nabla_{d]} X^a$, and the Ricci tensor is defined as $R_{ab} = {R^c}_{acb}$. For a world-line with proper time $\tau$ and 4-velocity $u$, we use the notation $D/\dif\tau = u^a \nabla_a$. The convention used for the wedge and symmetrized tensor products is such that $\dif x^a \wedge \dif x^b = 2 \dif x^{[a} \otimes \dif x^{b]}$ and $\dif x^a \otimes_\mathrm{s} \dif x^b = \dif x^{(a} \otimes \dif x^{b)}$, respectively. Total and partial derivatives with respect to coordinate time are denoted by $\dif f/\dif t$ and $\dot f = \partial_t f$, respectively. For a weak gravitational wave propagating in the $z$ direction, indices $I, J, K, L$ will be used to denote the transverse spatial coordinates, namely $(x^I) = (x,y)$. Units for which the speed of light in vacuum is unitary are used.

\section{General gravitational field} \label{sec:general-field}

\subsection{Cloud of freely falling test particles}

Consider spacetime to be modeled by a Lorentzian manifold $(\mathcal{M}^4, g)$, with its Levi-Civita connection $\nabla$, whose metric $g$ satisfies covariant field equations. We shall refer to this framework as a metric theory of gravity.

Let us consider a cloud of freely falling---subject only to the influence of gravity---collisionless test particles,\footnote{One might think of the cloud representing the test masses of a laser interferometer experiment, the fluid elements of a gas, or the atoms of a resonant bar.} hence moving along geodesics. Then, their 4-velocity field $u$ satisfies
\begin{align} \label{eq:nabla-u_orthogonal-to-u}
u^a \nabla_a u_b = 0, &&\nabla_a u_b \, u^b = 0.
\end{align}
The expansion $\theta$, the shear $\sigma$, and the vorticity $\omega$---the kinematical tensors---of the cloud are defined by
\begin{align} \label{eq:kinematical-tensors_def}
\theta \coloneqq \nabla_a u^a, &&\sigma_{ab} \coloneqq \nabla_{(a} u_{b)} - \frac{\theta}{3} P_{ab}, &&\omega_{ab} \coloneqq \nabla_{[a} u_{b]},
\end{align}
respectively,\footnote{In the literature, the vorticity is sometimes defined as the negative of the one defined here.} where
\begin{equation} \label{eq:projector-orthogonal-to-u}
P_{ab} \coloneqq g_{ab} + u_a u_b,
\end{equation}
is the projector orthogonal to $u$. The kinematical tensors constitute the irreducible decomposition of $\nabla_a u_b$, namely
\begin{equation} \label{eq:nabla-u_decomposition}
\nabla_a u_b = \frac{\theta}{3} P_{ab} + \sigma_{ab} + \omega_{ab}.
\end{equation}

The physical significance of the kinematical tensors comes from the fact that they determine the rate of change of the relative separation of neighboring particles of the cloud as follows. First, the relative separation is quantified by a separation vector field $\xi$, characterized by
\begin{align}
[\xi, u] = 0, &&\xi_a u^a = 0.
\end{align}
Thus, it follows that $D\xi^a/\dif\tau = \xi^b \nabla_b u^a$, which together with~\eqref{eq:nabla-u_decomposition} leads to the expression~\eqref{eq:relative-velocity} for the relative velocity. One then sees that: the expansion $\theta$ amounts to an isotropic rate of change of the relative separation, namely a change in volume; the shear $\sigma$ corresponds to a volume-preserving deformation; and the vorticity $\omega$ amounts to a rotation of the cloud (cf., e.g., \S 2.5 of \cite{ellis1971relativistic, ellis2009republication}). 

Moreover, it follows that $\xi$ satisfies the equation of geodesic deviation
\begin{equation}
\frac{D^2 \xi^a}{\dif\tau^2} = -{R^a}_{cbd} u^c \xi^b u^d,
\end{equation}
which, in terms of the tidal tensor
\begin{equation} \label{eq:tidal-tensor_def}
K_{ab} \coloneqq -R_{acbd} u^c u^d,
\end{equation}
leads to the expression~\eqref{eq:relative-acceleration} for the relative acceleration between neighboring test particles. The algebraic symmetries of the Riemann tensor imply that $K_{ab} = K_{(ab)}$ and $K_{ab} u^b = 0$, so $K$ has only six independent components. Furthermore, if one  combines~\eqref{eq:nabla-u_decomposition} with the Ricci identities, an expression for the tidal tensor in terms of the kinematical tensors and their derivatives is obtained; that expression can be simplified using the equations in~\eqref{eq:nabla-u_orthogonal-to-u} together with their first covariant derivatives, yielding (cf. \S 2.3 of \cite{ehlers1961beitraege,ehlers1993contributions})
\begin{multline} \label{eq:tidal-tensor_in-terms-of_kinematical-tensors}
K_{ab} = \frac{1}{3} \left( \frac{\dif\theta}{\dif\tau} + \frac{\theta^2}{3} \right) P_{ab} + \frac{2\theta }{3} \sigma_{ab} + \frac{D\sigma_{ab}}{\dif\tau}\\
	+ \sigma_{ac} {\sigma_b}^c - \omega_{ac} {\omega_b}^c.
\end{multline}

\subsection{Polarization modes} \label{sec:polarization-modes}

Now, we decompose the tidal tensor to obtain the polarization modes of gravitational waves. To do so, consider a gravitational wave to which one can associate a definite spacetime propagation direction, for example vacuum GR plane waves, either linearized or exact (cf. \S 35.3 and \S 35.9 of \cite{misner1973freeman}). Then, its spatial propagation direction, as seen in the instantaneous rest frame of the test particles, is given by a unit vector $n$ orthogonal to $u$, namely\footnote{An arbitrary observer with 4-velocity $u'$ would see a spatial propagation direction $n'$ determined by restrictions analogous to those in \eqref{eq:spatial-propagation-direction}.}
\begin{align} \label{eq:spatial-propagation-direction}
n_a n^a = 1, &&n_a u^a = 0.
\end{align}
Using $n$ and its orthogonal projector
\begin{equation}
p_{ab} \coloneqq g_{ab} + u_a u_b - n_a n_b,
\end{equation}
one can decompose the tidal tensor into the following parts
\begin{subequations} \label{eq:polarization-modes-def}
\begin{align}
&\Kl \coloneqq K_{ab} n^a n^b,\\
&\Ks \coloneqq p^{ab} K_{ab},\\
&\Kv_a \coloneqq {p_a}^b K_{bc} n^c,\\
&\Ktt_{ab} \coloneqq \left( {p_a}^c {p_b}^d - \frac{1}{2} p_{ab} p^{cd} \right) K_{cd},
\end{align}
\end{subequations}
which we shall call respectively, the longitudinal mode, the transverse scalar mode\footnote{Also called breathing mode.}, the vector mode, and the transverse tensor mode. From their definition, the vector and the transverse tensor modes satisfy $\Kv_a u^a = \Kv_a n^a = 0$, $\Ktt_{ab} u^b = \Ktt_{ab} n^b = 0$, and $g^{ab} \Ktt_{ab} = 0$; thus, each of them has two independent components. The longitudinal and transverse scalar modes comprise one polarization mode each; while the vector mode and transverse tensor mode comprise two polarization modes apiece, one for each independent component. In terms of the polarization modes, the tidal tensor gives
\begin{equation}
K_{ab} = \frac{\Ks}{2} p_{ab} + \Ktt_{ab} + 2 \Kv_{(a} n_{b)} + \Kl n_a n_b.
\end{equation}

The physical effects of the polarization modes stem from the foregoing equation and from the equation of geodesic deviation~\eqref{eq:relative-acceleration}. A small cloud of test particles initially at rest tends to: elongate longitudinally due to $\Kl$; expand or compress transversally due to a positive or negative $\Ks$, respectively;
elongate and compress respectively along two mutually orthogonal longitudinal-transverse directions due to $\Kv$; and elongate and compress respectively along two mutually orthogonal transverse directions due to $\Ktt$. The foregoing can be readily seen if one uses an
orthonormal frame adapted to $u$ and $n$ (cf. \cite{szekeres1965gravitational}) or specific coordinate systems adapted to the wave (cf., e.g., \S 11.4.3 of \cite{will2018theory} or \cite{fortini1982fermi, rakhmanov2014fermi}). The modes defined here correspond with those defined in the seminal work~\cite{eardley1973gravitational} (cf. their \S V A) for linearized plane waves with null propagation as follows:
$\Kl$ corresponds to their $p_1$; the two components $\Kv_I$ correspond to their $p_2$ and $p_3$; two independent components $\Ktt_{IJ}$ correspond to their $p_4$ and $p_5$; and $\Ks$ corresponds to their $p_6$.

Not all the kinematical tensors contribute equally to the polarization modes. From the expression for the tidal tensor in terms of the kinematical tensors~\eqref{eq:tidal-tensor_in-terms-of_kinematical-tensors}, it follows that
\begin{subequations} \label{eq:polarization-modes_in-terms-of_kinematical-tensors}
\begin{align}
\Kl &= \frac{1}{3} \left( \frac{\dif\theta}{\dif\tau} + \frac{\theta^2}{3} \right)\nonumber\\
	&\quad + n^a n^b \left( \frac{2\theta}{3} \sigma_{ab} + \frac{D \sigma_{ab}}{\dif\tau} + \sigma_{ac} {\sigma_b}^c - \omega_{ac} {\omega_b}^c \right),\\
\Ks &= \frac{2}{3} \left( \frac{\dif\theta}{\dif\tau} + \frac{\theta^2}{3} \right)\nonumber\\
	&\quad + p^{ab} \left( \frac{2\theta}{3} \sigma_{ab} + \frac{D \sigma_{ab}}{\dif\tau} + \sigma_{ac} {\sigma_b}^c - \omega_{ac} {\omega_b}^c \right),\\
\Kv_a &= {p_a}^b n^c \left( \frac{2\theta}{3} \sigma_{bc} + \frac{D\sigma_{bc}}{\dif\tau} + \sigma_{bd} {\sigma_c}^d - \omega_{bd} {\omega_c}^d \right),\\
\Ktt_{ab} &= \left( {p_a}^c {p_b}^d - \frac{1}{2} p_{ab} p^{cd} \right)\nonumber\\
	&\quad \times \left( \frac{2\theta }{3} \sigma_{cd} + \frac{D\sigma_{cd}}{\dif\tau} + \sigma_{ce} {\sigma_d}^e - \omega_{ce} {\omega_d}^e \right).
\end{align}
\end{subequations}
To gain some insight on the foregoing expressions, it is instructive to analyze the contributions of each kinematical tensor when the others vanish. For example, for a cloud with vanishing shear and vorticity, only the longitudinal and transverse scalar modes survive, determined completely by the combination $\dif\theta / \dif\tau + \theta^2/3$; this reveals a contribution from the expansion to the transverse scalar mode, as expected, but also to the longitudinal mode.
On the other hand, for a cloud with vanishing expansion and vorticity, all the modes are nonvanishing and are determined by the following combination of the shear $D \sigma_{ab} / \dif\tau + \sigma_{ac} {\sigma_b}^c$; this shows a contribution from some of its transverse components to the transverse tensor mode and a contribution from its longitudinal component to the longitudinal mode, as anticipated, but also shows a contribution from some of its transverse components to the transverse scalar mode and a contribution from its longitudinal-transverse components to the vector mode. Lastly, a rigid cloud, namely one with vanishing expansion and shear, exhibits a similar behavior to that of the preceding example, with all the modes nonvanishing and determined by $\omega_{ac} {\omega_b}^c$.

Certainly both the kinematical tensors and the polarization modes describe the same physical phenomena: the relative motion of neighboring test particles. In fact, there are some similitudes in the information each of them conveys, but also some dissimilitudes. For example, the volume-preserving deformation due to the shear $\sigma$ resembles the elongation-compression effect resulting from the transverse-tensor mode $\Ktt$, while the volume change due to the expansion $\theta$ is analogous to the area change resulting from the transverse scalar mode $\Ks$. However, a rotation of the cloud described by the vorticity $\omega$ cannot be described by a single polarization mode, as discussed in the preceding paragraph. Evidently, the description of the relative motion of test particles can be more transparent either from the kinematical tensors or from the polarization modes depending on the specific type of motion. Thus, the kinematical tensors can be seen as a complementary way of analyzing the physical effects of a gravitational wave. Additionally, they provide a new interpretation of the polarization modes in terms of their effect on the kinematical tensors, by means of the expressions in \eqref{eq:polarization-modes_in-terms-of_kinematical-tensors}.

\section{Weak gravitational waves} \label{sec:weak-waves}

\subsection{Approximation order}

Far away from a radiating system, the gravitational field is expected to be weak. Then, to certain precision, the metric can be approximated as
\begin{align}
g_{ab} = \eta_{ab} + h_{ab}, && |h_{ab}| \ll 1,
\end{align}
with respect to an almost inertial coordinate system $(x^a) = (t, x^i)$, say the laboratory frame. It is worth recalling that this coordinate system is determined up to Lorentz transformations and infinitesimal coordinate transformations. From now on, we shall consider expressions linear in the perturbation $h_{ab}$, and indices will be understood to be raised and lowered with the Minkowski metric. Linearized expressions for the purely geometric objects can be obtained, and are readily found in the literature (cf., e.g., \S 18.1 of \cite{misner1973freeman}). In particular, the linearized Riemann tensor yields
\begin{equation} \label{eq:riemann_linearized}
R_{abcd} = -\partial_a \partial_{[c} h_{d]b} + \partial_b \partial_{[c} h_{d]a},
\end{equation}
which is invariant upon infinitesimal coordinate transformations. Within linearized theory, the 4-velocity of a cloud of test particles is conveniently written in terms of their coordinate velocity
\begin{equation} \label{eq:coordinate-velocity_def}
v^i \coloneqq \frac{\dif x^i}{\dif t}.
\end{equation}
If the particles are freely falling, one expects them to be moving slowly, $|v^i|^2 \ll 1$, in accordance with the weak field approximation. Concretely we shall consider that
\begin{align}
|v^i|^3 \ll |h_{ab}|, &&|h_{ab} v^i| \ll |h_{ab}|,
\end{align}
thus keeping terms up to order $O(|v^i|^2)$ and $O(h_{ab})$. This is simply a first approximation to the problem.	Physical situations where this approximation order applies
are those for which the initial velocities of the test
particles before the passage of the wave are small
compared with the speed of light.

Then, from the normalization of $u$ one obtains
\begin{equation} \label{eq:4-velocity_linearized}
u = \left( 1 + \frac{1}{2} v_k v^k + \frac{1}{2} h_{tt} \right) \partial_t + v^i \partial_i,
\end{equation}
and, consequently, its orthogonal projector~\eqref{eq:projector-orthogonal-to-u} gives
\begin{equation} \label{eq:projector-orthogonal-to-u_linearized}
P = v_k v^k \dif t \otimes \dif t - 2 v_j \dif t \otimes_\mathrm{s} \dif x^j + \left( \eta_{ij} + h_{ij} + v_i v_j \right) \dif x^i \otimes \dif x^j,
\end{equation}
where $\otimes_\mathrm{s}$ was used to denote the symmetrized tensor product. Moreover, if the test particles fall freely, the geodesic equation implies
\begin{equation} \label{eq:coordinate-acceleration_general-free-fall}
\frac{\dif v^i}{\dif t} = -\dot{h}_{ti} + \frac{1}{2} \partial_i h_{tt},
\end{equation}
where $\dot{h}_{ab} = \partial_t h_{ab}$. Incidentally, the foregoing expression reduces to that for the Newtonian limit if the gravitational field is additionally quasi-stationary, namely if $\dot{h}_{ab}$ is negligible. For subsequent calculations, it is worth recalling the elementary chain rule for derivatives with respect to coordinate time, for example the coordinate acceleration yields $\dif v^i/\dif t = \dot{v}^i + v^k \partial_k v^i$ [cf. \eqref{eq:coordinate-velocity_def}].

\subsection{Linearized kinematical tensors and polarization modes}

Using the definition of the kinematical tensors~\eqref{eq:kinematical-tensors_def} together with~\eqref{eq:4-velocity_linearized}--\eqref{eq:coordinate-acceleration_general-free-fall}, a long but straightforward calculation leads to
\begin{subequations} \label{eq:linearized-kinematical-tensors}
\begin{align}
&\theta = \theta^\mathrm{M} + \frac{1}{2} \dot{h}^k{}_k,\\
&\sigma = \sigma^\mathrm{M} + \frac{1}{2} \left( \dot{h}_{ij} - \frac{1}{3} \dot{h}^k{}_k \eta_{ij} \right) \dif x^i \otimes \dif x^j,\\
&\omega = \omega^\mathrm{M} + \frac{1}{2} \partial_{[i} h_{j]t} \dif x^i \wedge \dif x^j,
\end{align}
\end{subequations}
where
\begin{subequations} \label{eq:Minkowskian-kinematical-tensors}
\begin{align}
\theta^\mathrm{M} &= \partial_k v^k,\\
\sigma^\mathrm{M} &= 2 \left( -v^k \partial_{(j} v_{k)} + \frac{1}{3} \partial_k v^k v_j \right) \dif t \otimes_\mathrm{s} \dif x^j\nonumber\\
	&\quad + \left( \partial_{(i} v_{j)} - \frac{1}{3} \partial_k v^k \, \eta_{ij} \right) \dif x^i \otimes \dif x^j,\\
\omega^\mathrm{M} &= v^k \partial_{[j} v_{k]} \dif t \wedge \dif x^j + \frac{1}{2} \partial_{[i} v_{j]} \dif x^i \wedge \dif x^j,
\end{align}
\end{subequations}
are purely kinematic contributions, in the sense that do not involve explicitly the gravitational field, present already in Minkowski spacetime.\footnote{Despite the fact that these purely kinematical quantities do not exhibit
explicitly the perturbation $h_{ab}$, they should not be
regarded as independent of it, since the freely falling condition
\eqref{eq:coordinate-acceleration_general-free-fall} in fact determines the velocities, and therefore those
quantities, in terms of the perturbation.}
The residual terms in~\eqref{eq:linearized-kinematical-tensors}, do involve explicitly the metric perturbation, and contribute differently for gravitational waves of different metric theories, as will be shown in \S \ref{sec:examples}. It is worth recalling that, for a given cloud of test particles, the kinematical
tensors are uniquely defined by covariant equations [cf. \eqref{eq:kinematical-tensors_def}--\eqref{eq:projector-orthogonal-to-u}] and are therefore well-defined independently of the specific coordinate system one uses; thus, $\theta$, $\sigma$, and $\omega$ are in particular gauge-invariant within the weak field approximation.

The reader might wonder if one could use a coordinate system in which the purely kinematical contributions \eqref{eq:Minkowskian-kinematical-tensors} simplify. In fact, since the particles are falling freely, one can use a Fermi normal coordinate system \cite{manasse1963fermi, fortini1982fermi, rakhmanov2014fermi} adapted to the worldline of one reference particle which thus will remain at rest with respect to this coordinate system. This would simplify slightly the expressions in \eqref{eq:Minkowskian-kinematical-tensors}. However, we prefer not to restrict the coordinate system this way and rather restrict it by gauge conditions that simplify the form of the perturbation for the gravitational wave, as will be shown in the examples of \S \ref{sec:examples}.

Moreover, from~\eqref{eq:riemann_linearized} and~\eqref{eq:4-velocity_linearized} it follows that the tidal tensor is
\begin{equation} \label{eq:tidal-tensor_linearized}
K_{ab} = \frac{1}{2} \left( \ddot{h}_{ab} + \partial_a \partial_b h_{tt} \right) - \partial_{(a} \dot{h}_{b)t},
\end{equation}
from which it follows that
\begin{equation}
K_{ab} = {\delta^i}_a {\delta^j}_b K_{ij}.
\end{equation}
In fact, a lengthy but direct calculation reveals that substituting the expressions for the linearized kinematical tensors \eqref{eq:linearized-kinematical-tensors}--\eqref{eq:Minkowskian-kinematical-tensors} into \eqref{eq:tidal-tensor_in-terms-of_kinematical-tensors}, while taking into account the freely falling condition \eqref{eq:coordinate-acceleration_general-free-fall}, yields an expression for $K_{ab}$ where all the terms involving the velocities of the particles cancel out, obtaining exactly the same expression as in \eqref{eq:tidal-tensor_linearized}.

For a wave propagating in the $+z$ direction, the tidal tensor decomposition of \S \ref{sec:polarization-modes} is made with respect to ${n^a = (P_{zz})^{-1/2} \, {P^a}_z}$, which gives
\begin{equation}
n = ( v_z + h_{tz} ) \partial_t + v_z v^I \partial_I + \left( 1 + \frac{1}{2} v_z^2 - \frac{1}{2} h_{zz} \right) \partial_z,
\end{equation}
while its orthogonal projector gives
\begin{equation}
p = v_I v^I \dif t \otimes \dif t - 2 v_J \dif t \otimes_\mathrm{s} \dif x^J + \left( \eta_{IJ} + h_{IJ} + v_I v_J \right) \dif x^I \otimes \dif x^J,
\end{equation}
where uppercase indices were used to denote spatial coordinates transverse to the spatial propagation direction of the wave, namely $(x^I) = (x, y)$. Then, the only nonvanishing components of the polarization modes~\eqref{eq:polarization-modes-def} yield simply
\begin{subequations}
\begin{align}
&\Kl = K_{zz},\\
&\Ks = {K^I}_I,\\
&\Kv_J = K_{zJ},\\
&\Ktt_{IJ} = K_{IJ} - \frac{{K^L}_L}{2} \eta_{IJ}.
\end{align}
\end{subequations}
These expressions coincide with those reported in the literature [cf., e.g., equation (6) of \cite{hyun2019exact}]. Explicitly in terms of the perturbation one obtains
\begin{subequations}
\begin{align} \label{eq:linearized-polarization-modes}
\Kl &= \frac{1}{2} \left( \ddot{h}_{zz} + \partial_z^2 h_{tt} \right) - \partial_z \dot{h}_{zt},\\
\Ks &= \frac{1}{2} \left( \ddot{h}^I{}_I + \partial^I \partial_I h_{tt} \right) - \partial^I \dot{h}_{It},\\
\Kv_I &= \frac{1}{2} \left( \ddot{h}_{Iz} + \partial_I \partial_z h_{tt} - \partial_I \dot{h}_{zt} - \partial_z \dot{h}_{It} \right),\\
\Ktt_{IJ} &= \frac{1}{2} \left( \ddot{h}_{IJ} - \frac{1}{2} \ddot{h}^K{}_K \eta_{IJ} \right) + \frac{1}{2} \left( \partial_I \partial_J h_{tt} - \frac{1}{2} \partial^K \partial_K h_{tt} \eta_{IJ} \right) \nonumber\\
	&\quad - \left( \partial_{(I} \dot{h}_{J)t} - \frac{1}{2} \partial^K \dot{h}_{Kt} \eta_{IJ} \right).
\end{align}
\end{subequations}

\section{Metric theories linearized waves} \label{sec:examples}

\subsection{General relativity} \label{sec:GR-example}

We shall evaluate the expressions for the kinematical tensors and the polarization modes obtained above on linearized gravitational waves of some metric theories. Our benchmark will be vacuum general relativity, so we will analyze it first. It is determined by the Einstein-Hilbert action
\begin{equation}
S [g_{ab}] = \frac{1}{16\pi G} \int \dif^4 x \sqrt{-g} \, R.
\end{equation}
For a linearized field, coordinate freedom allows one to impose, without loss of generality, the \emph{de Donder} gauge\footnote{Many other names are found in the literature for this gauge, such as: Harmonic, Hilbert, Einstein, Fock, Lorenz, and Lorentz.}
\begin{equation} \label{eq:de-donder}
\partial^a \left( h_{ab} - \frac{1}{2} h \, \eta_{ab} \right) = 0,
\end{equation}
where $h \coloneqq  {h^c}_c$. Then, the linearized field equations yield
\begin{eqnarray} \label{eq:linearized-EFE_wave-eq}
\Box_\eta h_{ab} = 0,
\end{eqnarray}
where $\Box_\eta = \partial^c \partial_c$. Moreover, the residual coordinate freedom within the de Donder gauge allows one to impose, also without loss of generality, that (cf. \S 4.4b of \cite{wald1984general} or \S 35.4 of \cite{misner1973freeman})
\begin{align} \label{eq:traceless-and-spatial-conditions_TT}
h = 0, &&h_{at} = 0.
\end{align}
Taken together, \eqref{eq:de-donder} and \eqref{eq:traceless-and-spatial-conditions_TT} define the transverse-traceless (TT) gauge, $h_{ab} = h^\mathrm{TT}_{ab}$. This perturbation corresponds to a massless spin-2 field (cf. \S III of \cite{fierz1939relativistic}). Note that, despite the similarity in the names and the labels of $h^\mathrm{TT}_{ab}$ and $\Ktt_{ab}$, they are related to different type of decompositions of symmetric tensors: the former involves differential relations~\eqref{eq:de-donder}, while the latter is defined in a purely algebraic manner~\eqref{eq:polarization-modes-def}. A thorough discussion on this matter can be found in \cite{ashtekar2017ambiguity}.

The TT gauge is determined up to Lorentz transformations, which allows one to write a monochromatic plane wave as
\begin{equation}
\label{eq:gr-plane-wave}
h^\mathrm{TT}_{ab} = h_+ e^+_{ab} + h_\times e^\times_{ab},
\end{equation}
where
\begin{align} \label{eq:tt-polarization-tensors}
e^+ = \dif x \otimes \dif x - \dif y \otimes \dif y, &&e^\times = 2 \dif x \otimes_\mathrm{s} \dif y,
\end{align}
and
\begin{align} \label{eq:hplus-and-htimes_plane-waves}
h_+ = \real \left( A_+ \e^{-\ii \Omega (t-z)} \right), &&h_\times = \real \left( A_\times \e^{-\ii \Omega (t-z)} \right),
\end{align}
for $A_+, A_\times \in \CC$, and $\Omega \in \RR$, constants.

Note that for this gravitational wave, particles that are stationary in the TT frame satisfy the freely falling condition~\eqref{eq:coordinate-acceleration_general-free-fall}; this is characteristic of GR. One then obtains the following kinematical tensors
\begin{align}
&\theta = \theta^\mathrm{M}, &&\sigma = \sigma^\mathrm{M} + \frac{1}{2} \dot{h}^\mathrm{TT}_{IJ} \dif x^I \otimes \dif x^J, &&\omega = \omega^\mathrm{M},
\end{align}
where $\dot{h}^\mathrm{TT}_{IJ} = \dot{h}_+ e^+_{IJ} + \dot{h}_\times e^\times_{IJ}$; while the polarization modes give
\begin{align}
\Ktt_{IJ} = \frac{1}{2} \ddot{h}^\mathrm{TT}_{IJ},
\end{align}
and $\Kl = 0, \Ks = 0, \Kv_a = 0$, where $\ddot{h}^\mathrm{TT}_{IJ} = \ddot{h}_+ e^+_{IJ} + \ddot{h}_\times e^\times_{IJ}$. One sees that $h^\mathrm{TT}_{IJ}$ contributes only to the transverse components of the shear and to the transverse tensor mode.

\subsection{Scalar curvature dependent Lagrangian} \label{sec:f-of-R}

Next, we consider an $f(R)$ theory, namely one determined by a Lagrangian that is a function of the scalar curvature $R$
\begin{equation}
S_f [g_{ab}] = \frac{1}{16\pi G} \int \dif^4 x \sqrt{-g} \, f(R).
\end{equation}
In particular we will consider a class of $f(R)$ theories subject to the following restrictions: that admits Minkowski as a solution in the absence of matter, which via the trace of the field equations requires $f(0) = 0$ (cf., e.g., \S II A of \cite{sotiriou2010f}); that is analytic at $R = 0$; and that satisfies $f'(0) / f''(0) > 0$. The first restriction is compulsory if one is to consider weak gravitational waves propagating on Minkowski spacetime, while the last restriction is necessary for causal wave propagation [cf.~\eqref{eq:f-of-R_perturbation-varphi-and-mu}--\eqref{eq:linearized-f-of-R_wave-eqs} below] and to include GR as a special case.

From the assumed analyticity of $f(R)$, it follows that only terms up to $O(R^2)$ in the action will contribute to the linearized field equations. Then, two different $f(R)$ will yield the same linearized field equations if their corresponding values of $f'(0)$ and $f''(0)$ coincide (cf., e.g., \S II of \cite{csenturk2012energy}). In particular, those for which $f''(0)$ vanishes yield the same vacuum linearized equations as those of GR, and thus to the gravitational waves of \S \ref{sec:GR-example}. Thus, let us now turn our attention to the nonvanishing $f''(0)$ case.

For a general linearized field, it is not possible to impose the very same gauge as that of GR, in particular the conditions in~\eqref{eq:traceless-and-spatial-conditions_TT}.
However, one can impose a gauge such that the perturbation is decomposed as (cf. \S III and \S IV of \cite{berry2011linearized})
\begin{align} \label{eq:f-of-R_perturbation-varphi-and-mu}
h_{ab} = \tilde{h}^\mathrm{TT}_{ab} + 2 \varphi \eta_{ab}, &&\varphi = -\frac{R}{6\mu^2}, &&\mu \coloneqq \sqrt{ \frac{f'(0)}{ 3f''(0) } },
\end{align}
with $\tilde{h}^\mathrm{TT}_{ab}$ subject to the transverse-traceless conditions~\eqref{eq:de-donder} and~\eqref{eq:traceless-and-spatial-conditions_TT}, yielding the linearized field equations
\begin{align} \label{eq:linearized-f-of-R_wave-eqs}
\Box_\eta \tilde{h}^\mathrm{TT}_{ab} = 0, &&\Box_\eta \varphi = \mu^2 \varphi.
\end{align}
One sees that the metric perturbation consists of a massless spin-2 field $\tilde{h}^\mathrm{TT}_{ab}$, which is the same as that of GR since it is subject to identical equations, and a massive spin-0 field $\varphi$. 

Consider monochromatic plane wave solutions for $\tilde{h}^\mathrm{TT}_{ab}$ and $\varphi$. Assuming both of them have the same wave propagation direction, allows one to write $\tilde{h}^\mathrm{TT}_{ab}$ exactly as the GR 
wave, namely
\begin{equation}
\label{eq:gr-like-plane-wave}
\tilde{h}^\mathrm{TT}_{ab} = h_+ e^+_{ab} + h_\times e^\times_{ab},
\end{equation}
supplemented with \eqref{eq:tt-polarization-tensors} and \eqref{eq:hplus-and-htimes_plane-waves}, and
\begin{align}
\varphi = \real \left( \alpha \e^{-\ii(\varpi t - qz)} \right), &&\varpi^2-q^2 = \mu^2, 
\end{align}
for $\alpha \in \CC$, and $\varpi, q \in \RR$, constants.

For this $f(R)$ gravitational wave, stationary particles in this ``generalized'' TT frame do not fall freely, since the right hand side of~\eqref{eq:coordinate-acceleration_general-free-fall} does not vanish, in contrast with GR. Then, for freely falling particles, the expressions in \eqref{eq:linearized-kinematical-tensors} lead to the following kinematical tensors
\begin{align}
\theta = \theta^\mathrm{M} + 3 \dot{\varphi}, &&\sigma = \sigma^\mathrm{M} + \frac{1}{2} \dot{\tilde h}^\mathrm{TT}_{IJ} \dif x^I \otimes \dif x^J, &&\omega = \omega^\mathrm{M},
\end{align}
while the polarization modes give
\begin{align}
\Kl = -\mu^2 \varphi, &&\Ks = 2\ddot{\varphi}, &&\Ktt_{IJ} = \frac{1}{2} \ddot{\tilde h}^\mathrm{TT}_{IJ},
\end{align}
and $\Kv_a = 0$. In addition to the contribution from $\tilde{h}^\mathrm{TT}_{IJ}$, which certainly is the same as that of GR, one observes that $\varphi$ contributes to the expansion, and to the longitudinal and transverse scalar modes (cf. table~\ref{tab:perturbation-fields_contributions}).

\begin{table}
	\centering
	\caption{Fields in the metric perturbation corresponding to linearized waves of the metric theories considered in \S \ref{sec:examples} and their contribution to both the kinematical tensors and the polarization modes are shown. The field $\tilde{h}^\mathrm{TT}_{ab}$ is the spin-2 massless field of GR, the field $\varphi$ is the spin-0 massive field arising in $f(R)$ theories (cf. \S \ref{sec:f-of-R}), and the field $\psi_{ab}$ is the spin-2 massive field appearing in Einstein-Bach gravity (cf. \S \ref{sec:einstein-bach}).}
	\label{tab:perturbation-fields_contributions}
	\begin{tabular}{cccccccc}
	\toprule
	Field	&&Spin	&&Mass	&&Contributes to\\
	\midrule
	$\tilde{h}^\mathrm{TT}_{ab}$	&&2	&&$=0$	&&$\sigma_{IJ}$, \ $\Ktt_{IJ}$\\
	$\varphi$	&&0	&&$\neq 0$	&&$\theta$, \ $\Kl$, \ $\Ks$\\
	$\psi^\mathrm{TT}_{IJ}$	&&	&&	&&$\sigma_{IJ}$, \ $\Ktt_{IJ}$\\
	$\psi_{zz}$	&&2	&&$\neq 0$	&&$\theta$, \ $\Kl$, \ $\Ks$, \ $\sigma_{ii}$\\
	$\psi_{zJ}$	&&	&&	&&$\sigma_{zJ}$, \ $\omega_{zJ}$, \ $\Kv_J$\\
	\bottomrule
	\end{tabular}
\end{table}

\subsection{Einstein-Bach gravity} \label{sec:einstein-bach}

Finally, let us analyze Einstein-Bach gravity\footnote{Also called Einstein-Weyl gravity.}, determined by the action
\begin{equation}
S_\gamma [g_{ab}] = \frac{1}{16\pi G} \int \dif^4 x \sqrt{-g} \left( R - \gamma C_{abcd} C^{abcd} \right),
\end{equation}
with $C_{abcd}$ the Weyl tensor and $\gamma$ a constant. The contribution of the Weyl squared term to the field equations is the Bach tensor~\cite{bach1921weylschen}, which is traceless. Thus, the trace of the vacuum field equations yields $R = 0$ in general. This theory is in fact a particular class of the theories whose Lagrangians are at most quadratic in the Riemann tensor~\cite{stelle1978classical}. We will assume $\gamma > 0$ for causal wave propagation [cf.~\eqref{eq:linearized-Einstein-Bach_wave-eqs} below].

For a general linearized field, $R = 0$ allows one to impose a gauge such that
\begin{equation}
h_{ab} = \tilde{h}^\mathrm{TT}_{ab} + \psi_{ab},
\end{equation}
with $\tilde{h}^\mathrm{TT}_{ab}$ subject to the transverse-traceless conditions~\eqref{eq:de-donder} and~\eqref{eq:traceless-and-spatial-conditions_TT}, and $\psi_{ab}$ traceless and subject to the de Donder condition~\eqref{eq:de-donder}, yielding the following linearized field equations (cf. \S II C 2 and \S IX A of \cite{bueno2017aspects}, or \S IV of \cite{alves2025gauges})
\begin{align} \label{eq:linearized-Einstein-Bach_wave-eqs}
\Box_\eta \tilde{h}^\mathrm{TT}_{ab} = 0, &&\Box_\eta \psi_{ab} = m^2 \psi_{ab}, &&m \coloneqq \sqrt{\frac{1}{2\gamma}}.
\end{align}
The metric perturbation now consists of a massless spin-2 field $\tilde{h}^\mathrm{TT}_{ab}$, which is also the same as that of GR, and a massive spin-2 field $\psi_{ab}$ (cf. \S I of \cite{fierz1939relativistic}). The latter has five independent components in general. For other commonly used gauge, see \cite{teyssandier1989linearised} and \S III of \cite{alves2025gauges}.

Consider monochromatic plane wave solutions for $\tilde{h}^\mathrm{TT}_{ab}$ and $\psi_{ab}$. Assuming both of them have the same wave propagation direction, allows one to write again $\tilde{h}^\mathrm{TT}_{ab}$ exactly as the GR wave [cf.~\eqref{eq:gr-like-plane-wave} supplemented with \eqref{eq:tt-polarization-tensors} and \eqref{eq:hplus-and-htimes_plane-waves}] and
\begin{subequations}
\begin{align}
&\psi_+ = \real \left( B_+ \e^{-\ii(\varpi t - qz)} \right),\\
&\psi_\times = \real \left( B_\times \e^{-\ii(\varpi t - qz)} \right),\\
&\psi_{zj} = \real \left( B_{j} \e^{-\ii(\varpi t - qz)} \right),\\
&\varpi^2 - q^2 = m^2,
\end{align}
\end{subequations}
for $B_+, B_\times, B_j \in \CC$, and $\varpi, q \in \RR$, constants, and
\begin{subequations}
\begin{align}
&\psi_{tt} = \left( \frac{q}{\varpi} \right)^2 \psi_{zz},\\
&\psi_{tj} = -\frac{q}{\varpi} \psi_{zj},\\
&\psi_{IJ} = \psi^\mathrm{TT}_{IJ} - \frac{1}{2} \left( \frac{m}{\varpi} \right)^2 \psi_{zz} \eta_{IJ},\\
&\psi^\mathrm{TT}_{IJ} = \psi_+ e^+_{IJ} + \psi_\times e^\times_{IJ}.
\end{align}
\end{subequations}

As with the $f(R)$ case, this gravitational wave does not allow stationary particles in this ``generalized'' TT frame to fall freely [cf.~\eqref{eq:coordinate-acceleration_general-free-fall}]. The kinematical tensors \eqref{eq:linearized-kinematical-tensors} give
\begin{subequations}
\begin{align}
\theta &= \theta^\mathrm{M} + \frac{1}{2} \left(\frac{q}{\varpi}\right)^2 \dot{\psi}_{zz},\\
\sigma &= \sigma^\mathrm{M} + \frac{1}{2} \left( \dot{\tilde h}^\mathrm{TT}_{IJ} + \dot{\psi}^\mathrm{TT}_{IJ} \right) \dif x^I \otimes \dif x^J\nonumber\\
	&\quad + \frac{m^2+2\varpi^2}{12\varpi^2} \dot{\psi}_{zz} \left( -\eta_{IJ} \dif x^I \otimes \dif x^J + 2 \dif z \otimes \dif z \right)\nonumber\\
	&\quad + \dot{\psi}_{zJ} \dif z \otimes_\mathrm{s} \dif x^J,\\
\omega &= \omega^\mathrm{M} + \frac{1}{2} \left(\frac{q}{\varpi}\right)^2 \dot{\psi}_{zJ} \dif z \wedge \dif x^J,
\end{align}
\end{subequations}
where $\partial_z \psi_{ab} = -(q/\varpi) \dot\psi_{ab}$ was used, while the polarization modes yield
\begin{subequations}
\begin{align}
&\Kl = -\frac{1}{2} \frac{m^4}{\varpi^2} \psi_{zz},\\
&\Ks = \frac{m^2}{2} \psi_{zz},\\
&\Kv_J = -\frac{m^2}{2} \psi_{zJ},\\
&\Ktt_{IJ} = \frac{1}{2} \left( \ddot{\tilde h}^\mathrm{TT}_{IJ} + \ddot{\psi}^\mathrm{TT}_{IJ} \right).
\end{align}
\end{subequations}
In addition to the GR contribution from $\tilde{h}^\mathrm{TT}_{IJ}$, the massive spin-2 field $\psi_{ab}$ contributes to all the kinematical tensors and to all the polarization modes: its TT part has a contribution similar to that of GR; its longitudinal component contributes to the expansion, to the diagonal components of the shear, and to the longitudinal and transverse scalar modes; and its longitudinal-transverse components contribute to the longitudinal-transverse components of both the shear and the vorticity, and to the vector mode. It stands out that the contributions to the vorticity and to the vector mode are nonvanishing. These contributions are summarized in table~\ref{tab:perturbation-fields_contributions}.

\section{Discussion}

We have considered a cloud of freely falling test particles acted upon by a gravitational wave, and studied relations between the kinematical tensors of the cloud and the polarization modes of the wave, which, to our knowledge, had not been studied before. The polarization modes of a gravitational wave with a definite spacetime propagation direction were expressed in terms of the kinematical tensors of the cloud in \S \ref{sec:polarization-modes}. In addition to the foreseen relations (cf. \S \ref{sec:intro}), the obtained expressions revealed that: the expansion contributes to the longitudinal mode; while both the shear and the vorticity contribute to the transverse scalar mode via some of their transverse components, and to the vector mode through some of their longitudinal-transverse components.
This analysis was general, and only assumed a definite wave propagation direction, having the advantage of being applicable to a wide variety of gravitational wave phenomena. More concrete aspects of gravitational waves might be incorporated, serving to obtain more specific and relevant relations, which would be interesting to explore in the future.

Additionally, we have calculated the kinematical tensors and the polarization modes for a cloud of slowly moving test particles that fall freely in an arbitrary linearized field. Being interested in studying gravitational waves beyond GR made it necessary to consider nonstationary particles, since not all theories allow the particles to be stationary in the laboratory frame, as GR does [cf.~\eqref{eq:coordinate-acceleration_general-free-fall}]. Linearized gravitational waves of certain metric theories were shown in \S \ref{sec:examples} to exhibit most of the anticipated relations between the kinematical tensors and the polarization modes. 
The GR wave contributes to the transverse components of the shear and to the transverse tensor mode. The $f(R)$ wave contributes additionally to the expansion, and to the longitudinal and transverse scalar modes. Finally, the Einstein-Bach wave contributes additionally to all the kinematical tensors and to all the polarization modes, standing out the nonvanishing contributions to the vorticity and to the vector mode. The contributions of each of these waves is summarized in table~\ref{tab:perturbation-fields_contributions}.

The inclusion of the kinematical tensors to the analysis of gravitational waves put forward in this work, provides new insights into the interpretation of the polarization modes, and offers a fresh perspective on devising new ways to find deviations from GR, both at the theoretical and at the phenomenological level.

It is reasonable to inquire about the relevance of the considered metric theories of gravity. It turns out that their linearized gravitational waves are representative of those of the larger class of metric theories determined by a Lagrangian that is a general function of the Riemann tensor. That this is so, comes from the fact that to obtain the linearized field equations, terms in the Lagrangian up to second degree in the Riemann tensor are the only ones that contribute (cf. \S II of \cite{csenturk2012energy}), which can be chosen to be the quadratic term of an $f(R)$ and the Weyl squared scalar considered in \S \ref{sec:f-of-R} and \S \ref{sec:einstein-bach}, respectively. Moreover, the $f(R)$ example turns out to be representative of second-order scalar-tensor theories, since their gravitational waves are essentially the same (cf. \S 3 of \cite{hou2018polarizations}). Hence, these results are applicable to a wide variety of metric theories of gravity.

Our general expressions for the linearized kinematical tensors and the polarization modes were written in terms of the components of the metric perturbation, which facilitates their evaluation on linearized gravitational waves of metric theories of gravity found in the literature. Although it has the disadvantage of being written in terms of gauge-dependent variables, the gravitational waves of the examples we considered were expressed in a completely fixed gauge, thus excluding nonphysical coordinate effects. Rewriting our expressions in terms of gauge-independent metric potentials (cf., e.g., \S 2.2 of \cite{flanagan2005basics} or \S III of \cite{dong2024polarization}), especially for the kinematical tensors, would ease their interpretation for general weak gravitational waves, but this goes beyond the scope of this work.

\begin{description}[left=0pt] \small

\item[Acknowledgments] The authors gratefully acknowledge valuable suggestions from E. Ay{\'o}n-Beato, D. Flores-Alfonso, and G. Aldaz, worthwhile discussions with H. Quevedo, as well as insightful inquiries from an anonymous referee that improved this paper. Some of the calculations done in this work were validated using \emph{Mathematica}'s package \mbox{\emph{xTensor}}~\cite{martin2008xtensor}.

\item[Funding] C.~M. acknowledges support from \href{https://ror.org/059ex5q34}{Secihti} graduate fellowship. P.~A.~S. work was supported by \href{https://ror.org/059ex5q34}{Secihti} postdoctoral fellowship, contract I1200/320/2022 Add. 2024.

\item[Data Availability Statement] Not applicable.

\item[Code Availability Statement] Not applicable.

\end{description}

\appendix

\section{Polarization modes in terms of the Newman-Penrose curvature scalars} \label{app:polarization-modes-NP}

To ease comparison of our results with others reported in the literature, we write here the polarization modes in terms of the irreducible parts of the Riemann tensor. Using the irreducible decomposition of the Riemann tensor (cf., e.g., \S 3.5 of \cite{stephani2003exact}) on \eqref{eq:polarization-modes-def}, supplemented with \eqref{eq:tidal-tensor_def}, allows one to write
\begin{subequations}
\begin{align}
\Kl &= \frac{R}{12} + \frac{1}{2} \left( S_{ab} n^a n^b - S_{ab} u^a u^b \right) - C_{abcd} n^a u^b n^c u^d,\\
\Ks &= \frac{R}{6} - \frac{1}{2} \left( S_{ab} n^a n^b + S_{ab} u^a u^b \right) + C_{abcd} n^a u^b n^c u^d,\\
\Kv_a &= \frac{1}{2} S_{bc} {p_a}^b n^c - C_{bcde} {p_a}^b u^c n^d u^e,\\
\Ktt_{ab} &= \frac{1}{2} S_{cd} {p_a}^c {p_b}^d + \frac{1}{4} \left( S_{cd} n^c n^d - S_{cd} u^c u^d \right) p_{ab}\nonumber\\
	&\quad - C_{cdef} {p_a}^c u^d {p_b}^e u^f - \frac{1}{2} C_{cdef} u^c n^d u^e n^f p_{ab},
\end{align}
\end{subequations}
where $S_{ab} = R_{ab} - (R/4) g_{ab}$ is the traceless Ricci tensor; and, recalling, $u$ is the 4-velocity field of the cloud of test particles, and $n$ is the spatial propagation direction, as seen in the instantaneous rest frame of the particles, of a non-necessarily weak gravitational wave [cf.~\eqref{eq:spatial-propagation-direction}]. It stands out that the scalar curvature $R$ does not contribute to the vector and transverse tensor modes.

One can obtain analogous expressions in terms of the Newman-Penrose curvature scalars~\cite{newman1962approach}. To accomplish this, introduce a complex null tetrad $(k,l,m,\cc{m})$, namely one such that $g_{ab} = -2 k_{(a} l_{b)} + 2 m_{(a} \cc{m}_{b)}$, defined by
\begin{align}
k^a = \frac{1}{\sqrt{2}} \left( u^a + n^a \right), &&l^a = \frac{1}{\sqrt{2}} (u^a - n^a), &&m_a u^a = 0.
\end{align}
Then, the polarization modes can be written as
\begin{subequations}
\begin{align}
&\Kl = \frac{R}{12} - S_{ab} m^a \cc{m}^b - (\Psi_2 + \cc{\Psi_2}),\\
&\Ks = \frac{R}{6} - \frac{1}{2} S_{ab} k^a k^b - \frac{1}{2} S_{ab} l^a l^b + \Psi_2 + \cc{\Psi_2},\\
&\Kv_a = \frac{1}{\sqrt{2}} \real \left( \left[ S_{bc} k^b \cc{m}^c - S_{bc} l^b \cc{m}^c + 2 \left( \cc{\Psi_1} - \Psi_3 \right) \right] m_a \right),\\
&\Ktt_{ab} = \real \left( \left[ S_{cd} \cc{m}^c \cc{m}^d - \left( \cc{\Psi_0} + \Psi_4 \right) \right] m_a m_b \right),
\end{align}
\end{subequations}
where $\Psi_0 = C_{abcd} k^a m^b k^c m^d$, $\Psi_1 = C_{abcd} k^a l^b k^c m^d$, $\Psi_2 = C_{abcd} k^a m^b \cc{m}^c l^d$, $\Psi_3 = C_{abcd} k^a l^b \cc{m}^c l^d$ and $\Psi_4 = C_{abcd} l^a \cc{m}^b l^c \cc{m}^d$. These relations were first reported for the particular case of linearized plane waves with null propagation in \cite{eardley1973gravitational} [cf. their (28) supplemented with their (10)--(12)]. The general expressions given in this appendix have also been reported in the literature (cf., e.g., \S III B of \cite{hyun2019exact}).

\bibliography{refs}

\end{document}